\documentclass[twocolumn]{aastex631}

\usepackage{comment}
\usepackage{savesym}
\savesymbol{tablenum}
\usepackage{siunitx}
\restoresymbol{SIX}{tablenum}
\DeclareSIUnit \parsec {pc}
\newcommand{\PULSARMINER}{\textsc{pulsar\_miner}}
\newcommand{\TEMPLATEBANKPEASOUP}{\textsc{3d/5d\_peasoup}}

\usepackage{threeparttable}

\begin{document}

\title{Missing for 20 years: MeerKAT re-detects the elusive binary pulsar M30B}

\author[0000-0003-3244-2711]{Vishnu Balakrishnan}
\affiliation{Max-Planck-Institut fur Radioastronomie, Auf dem Hügel 69, D-53121 Bonn, Germany}

\author[0000-0003-1307-9435]{Paulo C. C. Freire}
\affiliation{Max-Planck-Institut fur Radioastronomie, Auf dem Hügel 69, D-53121 Bonn, Germany}
\author[0000-0001-5799-9714]{S. M. Ransom}
\affiliation{National Radio Astronomy Observatory, 520 Edgemont Rd., Charlottesville, VA 22903, USA}
\author[0000-0001-6762-2638]{Alessandro Ridolfi}
\affiliation{INAF -- Osservatorio Astronomico di Cagliari, Via della Scienza 5, I-09047 Selargius (CA), Italy}
\affiliation{Max-Planck-Institut fur Radioastronomie, Auf dem Hügel 69, D-53121 Bonn, Germany}
\author[0000-0001-8715-9628]{E. D. Barr}
\affiliation{Max-Planck-Institut fur Radioastronomie, Auf dem Hügel 69, D-53121 Bonn, Germany}
\author[0000-0002-6089-7943]{W. Chen}
\affiliation{Max-Planck-Institut fur Radioastronomie, Auf dem Hügel 69, D-53121 Bonn, Germany}
\author[0000-0001-9518-9819]{Vivek~Venkatraman~Krishnan}
\affiliation{Max-Planck-Institut fur Radioastronomie, Auf dem Hügel 69, D-53121 Bonn, Germany}
\author[0000-0003-1361-7723]{D. Champion}
\affiliation{Max-Planck-Institut fur Radioastronomie, Auf dem Hügel 69, D-53121 Bonn, Germany}
\author[0000-0002-4175-2271]{M. Kramer}
\affiliation{Max-Planck-Institut fur Radioastronomie, Auf dem Hügel 69, D-53121 Bonn, Germany}
\affiliation{Jodrell Bank Centre for Astrophysics, Department of Physics and Astronomy, The University of Manchester, Manchester M13 9PL, UK}
\author[0000-0002-8396-2197]{T. Gautam}
\affiliation{Max-Planck-Institut fur Radioastronomie, Auf dem Hügel 69, D-53121 Bonn, Germany}
\author[0000-0001-5624-4635]{Prajwal V.~Padmanabh}
\affiliation{Max Planck Institute for Gravitational Physics (Albert Einstein Institute), D-30167 Hannover, Germany}
\affiliation{Leibniz Universit\"{a}t Hannover, D-30167 Hannover, Germany}
\affiliation{Max-Planck-Institut fur Radioastronomie, Auf dem Hügel 69, D-53121 Bonn, Germany}
\author[0000-0003-4137-4247]{Yunpeng Men}
\affiliation{Max-Planck-Institut fur Radioastronomie, Auf dem Hügel 69, D-53121 Bonn, Germany}

\author[0000-0002-9791-7661]{F. Abbate}
\affiliation{Max-Planck-Institut fur Radioastronomie, Auf dem Hügel 69, D-53121 Bonn, Germany}

%\collaboration{20}{(AAS Journals Data Editors)}

\author[0000-0001-9242-7041]{B. W. Stappers}
\affiliation{Jodrell Bank Centre for Astrophysics, Department of Physics and Astronomy, The University of Manchester, Manchester M13 9PL, UK}
\author[0000-0001-9784-8670]{I. Stairs}
\affiliation{Department of Physics and Astronomy, University of British Columbia, 6224 Agricultural Road, Vancouver, BC V6T 1Z1, Canada}
\author[0000-0001-9242-7041]{E. Keane}
\affiliation{6 School of Physics, Trinity College Dublin, University of Dublin, College Green, Dublin 2, D02 PN40, Ireland}
\author[0000-0001-5902-3731]{A. Possenti}
\affiliation{INAF -- Osservatorio Astronomico di Cagliari, Via della Scienza 5, I-09047 Selargius (CA), Italy}



%old abstract

\begin{abstract}
\end{abstract}

%New Abstract
\begin{abstract}
PSR~J2140$-$2311B is a 13-ms pulsar discovered in 2001 in a 7.8-hour Green Bank Telescope (GBT) observation of the core-collapsed globular cluster M30 and predicted to be in a highly eccentric binary orbit. This pulsar has eluded detection since then, therefore its precise orbital parameters have remained a mystery until now. In this work, we present the confirmation of this pulsar using observations taken with the UHF receivers of the MeerKAT telescope as part of the TRAPUM Large Survey Project.
Taking advantage of the beamforming capability of our backends, we have localized it, placing it \SI{1.2(1)} \arcmin\, from the cluster centre.
Our observations have enabled the determination of its orbit: it is highly eccentric ($e = 0.879$) with an orbital period of $6.2$ days. We also measured the rate of periastron advance, $\dot{\omega} = 0.078 \pm 0.002\, \rm deg \, yr^{-1}$. Assuming that this effect is fully relativistic, general relativity provides an estimate of the total mass of the system, $M_{\rm TOT} = 2.53 \pm 0.08$ M$_{\odot}$, consistent with the lightest double neutron star systems known. Combining this with the mass function of the system gives the pulsar and companion masses of $m_p < 1.43 \, \rm M_{\odot}$ and $m_c > 1.10 \, \rm M_{\odot}$ respectively. The massive, undetected companion could either be a massive WD or a NS. M30B likely formed as a result of a secondary exchange encounter. Future timing observations will allow the determination of a phase-coherent timing solution, vastly improving our uncertainty in $\dot{\omega}$ and likely enabling the detection of additional relativistic effects which will determine $m_p$ and $m_c$.  
\end{abstract}


%%%%%%%%%%%%%%%%%%%%%%%%%%%%
%Paper Structure

%Intro
% Why do we search in GC? 
% A brief history. Progress made by FAST and MKAT. 
%Mystery behind M30B

%Observations (L-BAND, UHF, TRAPUM + MeerTIME)
%Search Analysis (Pulsarminer + template-bank)
%Localisation (SeeKAT)
%Orbital constraints (polynomial method)
%Timing (TEMPO + DRACULA, measurement of omega dot)
%What do these orbital parameters mean in terms of evolution of M30B
%Future work

%%%%%%%%%%%%%%%%%%%%%%%%%%
\section{Introduction} \label{sec:intro}
Globular clusters (GCs) are dense spheroidal arrangement of stars held together by their own gravity. Their high stellar densities (10$^{3-6}\, \rm pc^{-3}$) enable dynamical interactions, where some of these old NSs --- which would have remained undetectable if located in the Galactic disk --- can gain a stellar mass companion, for example, a main-sequence (MS) star either in binary - single star encounters, or direct NS - MS encounters \citep{Verbunt1987,1995ApJS...99..609S,1995MNRAS.276..876D,1998MNRAS.301...15D}.

In the resulting binaries, a MS or giant star transfers mass and angular momentum to the NS.
During this accretion process, commonly referred to as `recycling', thermal X-ray emission is produced due to frictional heating from the in-falling matter, making these systems detectable as low, intermediate, or high-mass X-ray binaries depending on the mass of the donor star. In GCs, only low-mass stars are still on the main sequence, therefore any X-ray binaries containing NSs are low-mass X-ray binaries (LMXBs; \citealt{Clark1975}). Because low-mass MS companions evolve slowly, LMXBs are very long-lived. At the end of this process, a recycled radio pulsar (a `millisecond pulsar', or MSP, with $P < 20$ ms) emerges (e.g. \citealt{1982Natur.300..728A}).
These systems mostly resemble the MSPs in the Galactic disk, which have a wide variety of companions (black widows, redbacks, white dwarfs and some isolated MSPs as well); all of them produced by unperturbed stellar evolution; the majority have orbits with low ($e < 10^{-3}$) eccentricities \citep{2005AJ....129.1993M}\footnote{\url{https://www.atnf.csiro.au/research/pulsar/psrcat/}}.

These dynamical formation channels for LMXBs explain why LMXBs and MSPs are so abundant in GCs relative to the Galaxy. At the time of writing (2022 December), 272 radio pulsars in 38 GCs are known (see `GC Pulsar Catalog`\footnote{See \url{http://www.naic.edu/~pfreire/GCpsr.html} for an up-to date count.}) of which 245 ($\sim 90 \%$) are MSPs. Per unit stellar mass, GCs are estimated to have three orders of magnitude more LMXBs and MSPs than the Galactic disk \citep{Clark1975,vandenBerg2020}. Some of these channels, like direct collisions of NSs with MS stars \citep{1992ApJ...401..246D}, can also explain some of the exotic objects found in GCs, like ultra-compact X-ray binaries \citep{2005ApJ...621L.109I}, see \cite{2022ApJ...931...84Y} for a recent review. Furthermore, some dynamical interactions, like perturbations from nearby stars, explain why a significant percentage of the MSPs - white dwarf systems in GCs have mildly eccentric orbits \citep{1992RSPTA.341...39P,2005ASPC..328..147C,2008IAUS..246..291R}.

However, the extreme stellar densities at the cores of the GCs with the highest interaction rates per binary, $\gamma$ \citep{2014A&A...561A..11V} - especially the core-collapsed GCs - imply that stars are likely to go through repeated gravitational interactions with other stars and binaries in the core. This leads to the formation, in these GCs, of binary systems where already fully recycled MSPs acquire massive companions in additional exchange encounters \citep{1991ApJ...374L..41P,2004ApJ...606L..53F,2012ApJ...745..109L,2015ApJ...807L..23D,2021MNRAS.504.1407R,2022A&A...664A..27R, 2022ApJ...934L...1K}. If these massive companions are degenerate, so these orbits retain the high eccentricity of the systems after the exchange encounters. Such systems could even include MSP - black hole binaries (e.g. \citealt{2019ApJ...877..122Y}).

These systems are especially interesting because the rotational stability of MSPs, which rivals atomic clocks on long timescales (e.g. \citealt{2020MNRAS.491.5951H}) makes them extremely useful for a diverse array of applications: their orbital eccentricities and large companion masses enables precise mass measurements for the MSPs and their companions \citep{2012ApJ...745..109L,2019MNRAS.490.3860R} and, at least in one case so far, tests of gravity theories \citep{2006ApJ...644L.113J}. Even if they are not in eccentric binaries, MSPs in globular clusters can be used to probe their gravitational potentials \citep{2017MNRAS.471..857F,2017ApJ...845..148P,2017MNRAS.468.2114P, 2018MNRAS.481..627A}.

M30 (NGC 7099) is a GC located at a distance of 8.1 kpc from the sun, at Galactic coordinates $l = 27.18 ^\circ$, $b = -46.84 ^\circ$  (\citealt{1996yCat.7195....0H}, 2010 revision), and has an estimated age of 12.9 Gyr \citep{2010MNRAS.404.1203F}. This GC is of particular interest for pulsar searching as it is a core-collapsed cluster and has shown significant evidence of mass segregation \citep{Howell_2000}. Its core has a radius of \SI{3.6}{\arcsecond} and its half-light radius is \SI{61.8}{\arcsecond} \citep{Harris2010}.

Previous searches for pulsars in M30 using observations taken at the 100-m Green Bank Telescope (GBT), yielded the discovery of two radio pulsars: PSR~J2140$-$2310A (M30A), a 11.01-ms eclipsing pulsar in a 4.17-h orbit; and PSR~J2140$-$2311B (M30B) \citep{2004ApJ...604..328R}, a 13.0-ms binary pulsar. Based on the spin frequency evolution seen in the 7.8 h discovery observation in 2001, \citet{2004ApJ...604..328R} could infer that this pulsar must be in a highly eccentric ($e \geq 0.45$), relativistic orbit. However, a precise characterisation of this orbit was not possible because the pulsar was not detected in any other observations made with the GBT with the total time spent on source adding up to \SI{30}{h}. The reason for this is that it was discovered while its flux was being amplified by diffractive scintillation, which produces a strong modulation in the flux densities of the pulsars in this cluster (see Fig. 3 of \citealt{2004ApJ...604..328R}, which shows the flux density variations observed for M30A). Because of this, the basic characteristics of this system have remained a mystery for the last 20 years.

Aided by sensitivity gains offered by the MeerKAT telescope, we present the first set of detections of this pulsar since the discovery observation in 2001, which have revealed the nature of this system.

\section{Observations and data reduction}\label{sec:observation}
We observed M30 using MeerKAT, with at least 56 antennas per observation, on 9 occasions between 2020 December and 2022 September (see table \ref{tab:list_observations m30b} for details). Our initial campaign consisted of four observations, each lasting 60 minutes, taken as part of the TRansients And PUlsars with Meerkat
(TRAPUM\footnote{\url{http://www.trapum.org}}; \citealt{Stappers_Kramer2016}) Large Survey Project (LSP) between December 2020 and January 2021.

\begin{table*}
\begin{threeparttable}
\renewcommand{\arraystretch}{1.1}
\setlength{\tabcolsep}{0.15cm}
\footnotesize
\centering
\caption{List of MeerKAT observations of M30B recorded and analysed for this work. $t_{\rm samp}:$ sampling time, $N_{\rm pol}:$ Number of polarizations recorded, $f_{\rm c}:$ central frequency, $\Delta f:$ bandwidth, $N_{\rm ant:}$ number of MeerKAT antennas used, $N_{\rm beam}:$ number of tied-array beams recorded. All APSUSE observations were incoherently dedispersed to DM = \SI{25.06}{\parsec \per \cubic \centi \meter} and downsampled to 256 channel filterbank files. All PTUSE observations were recorded with full stokes information, coherently dedispersed to a DM of \SI{25.06}{\parsec \per \cubic \centi \meter} and downsampled to 256 channel psrfits files to save disk space.}
\label{tab:list_observations m30b}
\begin{tabular}{crrrrrcccccr}
\hline
Obs. id            & \multicolumn{1}{c}{Start Time}    & \multicolumn{1}{c}{Start Time}       & \multicolumn{1}{c}{Length}    & \multicolumn{1}{c}{Backend} & \multicolumn{1}{c}{$t_{\rm samp}$} & \multicolumn{1}{c}{$N_{\rm pol}$}& \multicolumn{1}{c}{$f_{\rm c}$}  & \multicolumn{1}{c}{$\Delta f$}  & \multicolumn{1}{c}{$N_{\rm chan}\phantom{*}$} & \multicolumn{1}{c}{$N_{\rm ant}$}  & $N_{\rm beam}$    \\
               & \multicolumn{1}{c}{(Date)}        & \multicolumn{1}{c}{(MJD)}            & \multicolumn{1}{c}{(s)}  & \multicolumn{1}{c}{}  & \multicolumn{1}{c}{($\mu$s)} &  & (MHz)        & (MHz) &               &   \\
\hline
01L       & 2021 Dec 17 & 59200.512 & 3600 & APSUSE & 76.56 & 1 & 1284 & 856 &  4096 & 56   & 287 \\
02L       & 2021 Dec 29 & 59212.572 &  3600 & APSUSE & 76.56 & 1 & 1284 & 856 &  4096 & 56   & 287 \\

03U       & 2021 Jan 22 & 59236.453 &  3600 & APSUSE & 60.24 & 1 &  816 & 544 &  4096 & 56   & 287 \\
04U       & 2021 Jan 30 & 59244.488 &  3600 & APSUSE & 60.24 & 1 &  816 & 544 &  4096 & 56   & 275 \\

05U\tnote{a}     & 2022 Jun 29 & 59759.837 &  3600 &  PTUSE &  9.41 & 4 &  816 & 544 & 1024 & 60 &   1 \\
06U\tnote{a}     & 2022 Jun 30 & 59760.179 &  3600 &  PTUSE &  9.41 & 4 &  816 & 544 & 1024 & 60 &   1 \\
07U\tnote{a}     & 2022 Jun 30 & 59760.880 &  3600 &  PTUSE &  9.41 & 4 &  816 & 544 & 1024 & 60 &   1 \\
08U\tnote{a}     & 2022 Jul 02 & 59762.877 &  3600 &  PTUSE &  9.41 & 4 &  816 & 544 & 1024 & 60 &   1 \\
09U     & 2022 Sep 09 & 59831.785 &  16200 &  PTUSE &  9.41 & 4 &  816 & 544 & 1024 & 64 &  1 \\

\hline
\end{tabular}
\begin{tablenotes}
\item[a] These observations are phase connected in our timing solution provided in table \ref{tab:timing_solution_M30B_DD.par}. The remaining observations have been fitted with arbitrary time-offsets in the form of `JUMP' statements.\\
\end{tablenotes}
\end{threeparttable}
\end{table*}

Of these, the first two observations were recorded with the L-band receivers, with a central frequency of \SI{1284}{\mega \hertz} and a bandwidth of \SI{856}{\mega \hertz} split into 4096 frequency channels and sampled every \SI{76.56} {\micro \second}. The next two observations were recorded with the Ultra High Frequency (UHF) receivers, centered at the frequency of \SI{816}{\mega \hertz} with a bandwidth of \SI{544}{\mega \hertz}, also split into 4096 channels, and sampled every \SI{60.24} {\micro \second}. We used the Filterbanking Beamformer User Supplied Equipment \citep[FBFUSE,][]{Barr2018} as the backend to form between 275-287 synthesized beams on the sky with an overlap fraction\footnote{The boundary of the synthesized beams in the tiling overlap each other with the power level equal to this ratio.} of 0.8, enabling arcsecond localization of pulsars after initial detection. The beam tiling pattern corresponding to each pointing was estimated based on an optimal hexagonal packing approach of elliptical beams. This was done using the  \textsc{Mosaic}\footnote{\url{https://github.com/wchenastro/Mosaic}} software \citep{2021JAI....1050013C}.
The beams were processed on-line by the Accelerated Pulsar Search User Supplied Equipment (APSUSE) computing cluster, where data from each beam are converted into an 8-bit, Stokes-I pulsar-search mode file, based on the \textsc{sigproc} \textsc{filterbank} format \citep{2011ascl.soft07016L}. We then incoherently dedispersed the observations off-line, using a dispersion measure (DM) of \SI{25.06}{\parsec \per \cubic \centi \meter} (which corresponds to the DM of the other known pulsar, M30A), and then downsampled the data in frequency by a factor of 16 (bringing the total number of frequency channels to 256) so as to reduce the data volume. 

Based on the results obtained from the TRAPUM observations, we also carried out a follow-up orbital campaign, with a pseudo-log cadence, between 2022 June and July . Each of these observations was 60 minutes, with the exception of a 280-minute (4.5 h) periastron-passage observation in September 2022. These follow-up observations were made using at least 60 antennas, the UHF receivers and the Pulsar Timing User Supplied Equipment (PTUSE) backend \citep{2020PASA...37...28B}. PTUSE recorded the data from a single tied-array beam placed at the nominal centre of M30.\footnote{Additionally, we also recorded TRAPUM search-mode observations in parallel to enable further pulsar and transient searches of the cluster.} The full bandwidth of \SI{544}{\mega \hertz} was split into 1024 frequency channels, coherently dedispersed to a DM of \SI{25.06}{\parsec \per \cubic \centi \meter}, downsampled to 256 frequency channels and saved in PSRFITS format \citep{2004PASA...21..302H}. PTUSE observations were also recorded in search-mode since at the time we did not yet have an accurate orbital ephemeris, and to allow further pulsar searches. These observations were recorded in full-Stokes mode to enable polarimetric measurements, and sampled every \SI{15.06} {\micro \second} for high-resolution pulsar timing.

All the observations used the Inter-Quartile Range Mitigation algorithm \citep{2022MNRAS.510.1393M} as a first-pass to filter out bright Radio-frequency interference (RFI) signals. The data were then shipped on hard drives to Garching, Germany, where the primary search analysis was conducted using the Max Planck Computing and Data Facility (MPCDF) Hercules\footnote{\url{https://docs.mpcdf.mpg.de/doc/computing/clusters/systems/Radioastronomy.html}} cluster.  

\subsection{Search Analysis}
Our search analysis consists of two pipelines implementing different search algorithms. First is \PULSARMINER \footnote{\url{https://github.com/alex88ridolfi/PULSAR_MINER}}, a user-friendly wrapper of  \textsc{presto}\footnote{\url{https://github.com/scottransom/presto}} which is a fourier-domain acceleration search pipeline sensitive to binary pulsars in constant acceleration ($P\mathrm{_{orb}}~\gtrsim$~10~$T\mathrm{_{obs}}$; \citealt{2002AJ....124.1788R, 2003ApJ...589..911R}). Our second pipeline called \TEMPLATEBANKPEASOUP \footnote{\url{https://github.com/vishnubk/5D_Peasoup}} uses the template-bank algorithm to search for pulsars in compact circular orbit binaries by coherently searching across three Keplerian parameters in the time-domain \citep{2022MNRAS.511.1265B}. 

Before commencing searching, we cleaned all of the observations again using \textsc{presto}'s \texttt{rfifind} program, which masks frequency channels and sub-integrations contaminated by radio frequency interference (RFI) using appropriate user-defined thresholds. Additionally, we removed Fourier frequencies identified in a topocentric zero-DM timeseries as they are almost certainly caused by RFI. We then dedispersed the data between 23-28~\SI{} {\parsec \per \cubic \centi \meter} and transformed our observation frame of reference to the solar-system barycenter by adding or subtracting appropriate delays from the dedispersed timeseries using the DE421 JPL Ephemerides \citep{2009IPNPR.178C...1F}. We searched for both isolated and binary pulsars using the \texttt{accelsearch} routine in \textsc{presto}. For our searches, we used the GPU version\footnote{\url{https://github.com/jintaoluo/presto_on_gpu}} of \textsc{presto} with a z$_{\rm max}$ = 1200 and performed incoherent harmonic summing of up-to 16 harmonics to be sensitive to narrow-duty cycle pulsars. 

%Old stuff
\begin{comment}

In the case of a binary pulsar, as the pulsar moves around its orbit, from the observer's point of view it creates a Doppler modulation of the apparent spin-frequency of the pulsar which smears the signal into adjacent Fourier frequency bins. Acceleration searches work under the assumption that this change can be approximated by a single parameter called ``acceleration'', which is assumed to be constant during the observation. This assumption is valid if the observation does not sample a large fraction of the orbit i.e $P\mathrm{_{orb}}~\geq$~10~$T\mathrm{_{obs}}$ \citep{2003ApJ...589..911R, 2015MNRAS.450.2922N}. Searches in acceleration can be conducted in both the time-domain \citep{1991ApJ...368..504J} and frequency-domain \citep{2002AJ....124.1788R}. In a frequency-domain search (for e.g. \textsc{presto}), acceleration is usually parameterized by a ``z'' value which is defined as the frequency derivative in units of Fourier bins (i.e. how many bins the signal drifts during the observation). For our searches, we used the GPU version\footnote{\url{https://github.com/jintaoluo/presto_on_gpu}} of \textsc{presto} with a z$_{\rm max}$ = 1200 and performed incoherent harmonic summing of up-to 16 harmonics to be sensitive to narrow-duty cycle pulsars. 
\end{comment}

Additionally, in order to be sensitive to binary pulsars in more compact orbits, we performed the search not only on the full 60-minute observation, but also split the data into 15- and 30-minute chunks and searched each segment with the same z$_{\rm max}$ value of 1200. This gives us extra sensitivity towards pulsars in compact orbits (i.e. P$_{\rm orb} \gtrsim$ $150 \, \rm min$ for the 15-min segments) at the cost of raising our minimum flux density limit as sensitivity improves with $T\mathrm{_{obs}}^{1/2}$. These searches also improve our chances for finding a pulsar which shows significant diffractive scintillation.

Finally, in order to take advantage of the sensitivity offered by the full 60-minute observation, but increase our sensitivity towards compact orbits (i.e not be limited to $P\mathrm{_{orb}}~\gtrsim$~10~$T\mathrm{_{obs}}$) we carried out a template-bank search \citep{2009PhRvD..79j4017M, 2011PhDT.......293K, 2013ApJ...774...93K, Allen_2013} where we expand from a one-dimensional acceleration search to a three-dimensional Keplerian parameter search (orbital period, projected semi major axis and initial orbital phase) assuming a circular orbit binary. Here we searched for orbits between 4 and 10\,h in the 60-minute observation and between  2 and 5\,h in the 30-minute chunks, with initial orbital phase between 0 and 2 $\pi$, mismatch of 10 \% and a coverage of 90 \% assuming a minimum pulsar spin-period of 2 ms, minimum pulsar mass of 1.4 M$_{\odot}$ and a maximum companion mass of 8 M$_{\odot}$. We refer the interested readers to \S 3 of \citet{2021MNRAS.504.1407R} and \S 2 of \citet{2022MNRAS.511.1265B} for a more in-depth review of the \PULSARMINER \, and the \TEMPLATEBANKPEASOUP\ pipelines respectively. On average, we folded approximately 350 pulsar candidates per beam for all our acceleration searches and 1000 pulsar candidates per beam for our template-bank searches\footnote{Larger numbers of candidates are expected for higher-order template-bank searches since our computational trials scale up due to the addition of more binary parameters.}. Multiplying this number by the total number of synthesized beams times the four initial observations taken under the \textsc{TRAPUM} project, approximately 1.5 million pulsar candidates were produced which was infeasible to inspect manually. Therefore we used machine-learning based pipelines to extract the most interesting pulsar candidates. We used the Pulsar Image Classification system (PICS; \citealt{2014ApJ...781..117Z}) and Pulsar Candidate Identification Using Semi-Supervised Generative Adversarial Networks (SGAN;  \citealt{2021MNRAS.505.1180B}) and manually inspected candidates which scored above a threshold of 0.5 using either of these algorithms. This reduced our candidate viewing load from 1.5 million to approximately 5000 ($\sim 0.3 \%$).  No new pulsars have been discovered in our analysis. However, our pipelines blindly re-detected and confirmed the elusive binary pulsar PSR~J2140$-$2311B described in the next section.

\section{Results}

\begin{figure}[ht!]
    %\centering
    \includegraphics[width=0.55\textwidth]{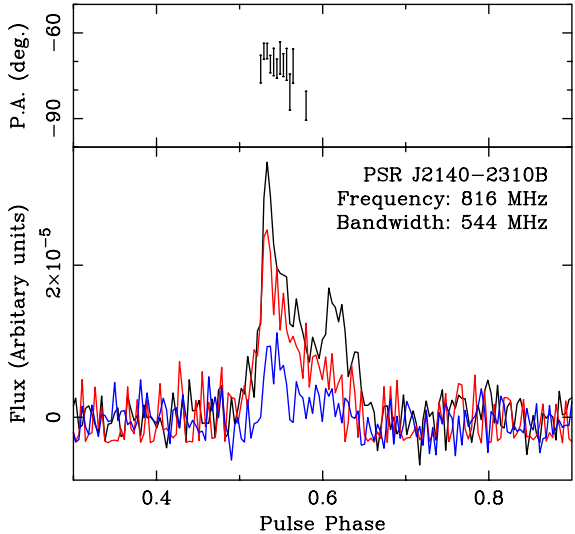}
    \caption{Bottom Panel: High S/N polarization pulse profile of M30B obtained by summing phase-connected observations between 05U-08U recorded in the UHF band. Black line represents the total intensity, red and blue lines display the linear and circular polarization pulse profile respectively. Top Panel: Measurements of Position Angle (PA) in degrees above four sigma of the linear polarization.}
    \label{fig:polarization profile}
\end{figure}

\subsection{Confirmation of PSR~J2140$-$2311B (M30B)}
We redetected M30B in two out of four of our initial set of TRAPUM observations. Both of our detections were from data recorded in the UHF band taken during 2021 January 22 and 30 (obs id: 03U and 04U) with a folded significance of 12.4 and 14.0 sigma respectively. The measured barycentric spin-period of 12.9896559(7) ms with a detected DM = \SI{25.07}{\parsec \per \cubic \centi \meter} in the Jan 22 observation immediately confirmed that this was indeed M30B initially reported in \citet{2004ApJ...604..328R}. The absence of any clear detection from searches in the L-band is consistent with the result of most GBT observations. 

M30B has been consistently detected in our follow-up orbital campaigns between 2022 June-September, which were recorded in the UHF band; this is likely caused by a combination of factors: the steep spectrum of the pulsar, the fact that scintles are narrower at these lower frequencies (thus leading to more of them being averaged within a given band), higher sensitivity of MeerKAT (Gain G = 2.63 K/Jy, assuming 60 antennas used) compared to GBT (G = 2.0 K/Jy) and the wider frequency bandwidth of the MeerKAT UHF receivers (Bandwidth: 544 MHz) compared to the GBT 820 MHz receiver (Bandwidth: 200 MHz). Using PTUSE full-Stokes observations taken between 2022 Jun - July (obs id: 05U-08U), we were able to obtain a high-S/N polarization pulse profile of this pulsar using standard routines in the \textsc{PSRCHIVE}\footnote{\url{https://psrchive.sourceforge.net/}} \citep{2012AR&T....9..237V} software suite. We detected a rotation measure (RM) = 8 $\pm$ 6 \, rad m$^{-2}$ which was used to de-Faraday the folded archive to produce a linear and circular polarization pulse profile. This is shown in Figure \ref{fig:polarization profile}.

\subsection{Localization}
\label{sec:localisation}
\begin{figure*}[ht!]
    \centering
    \includegraphics[width=0.65\textwidth]{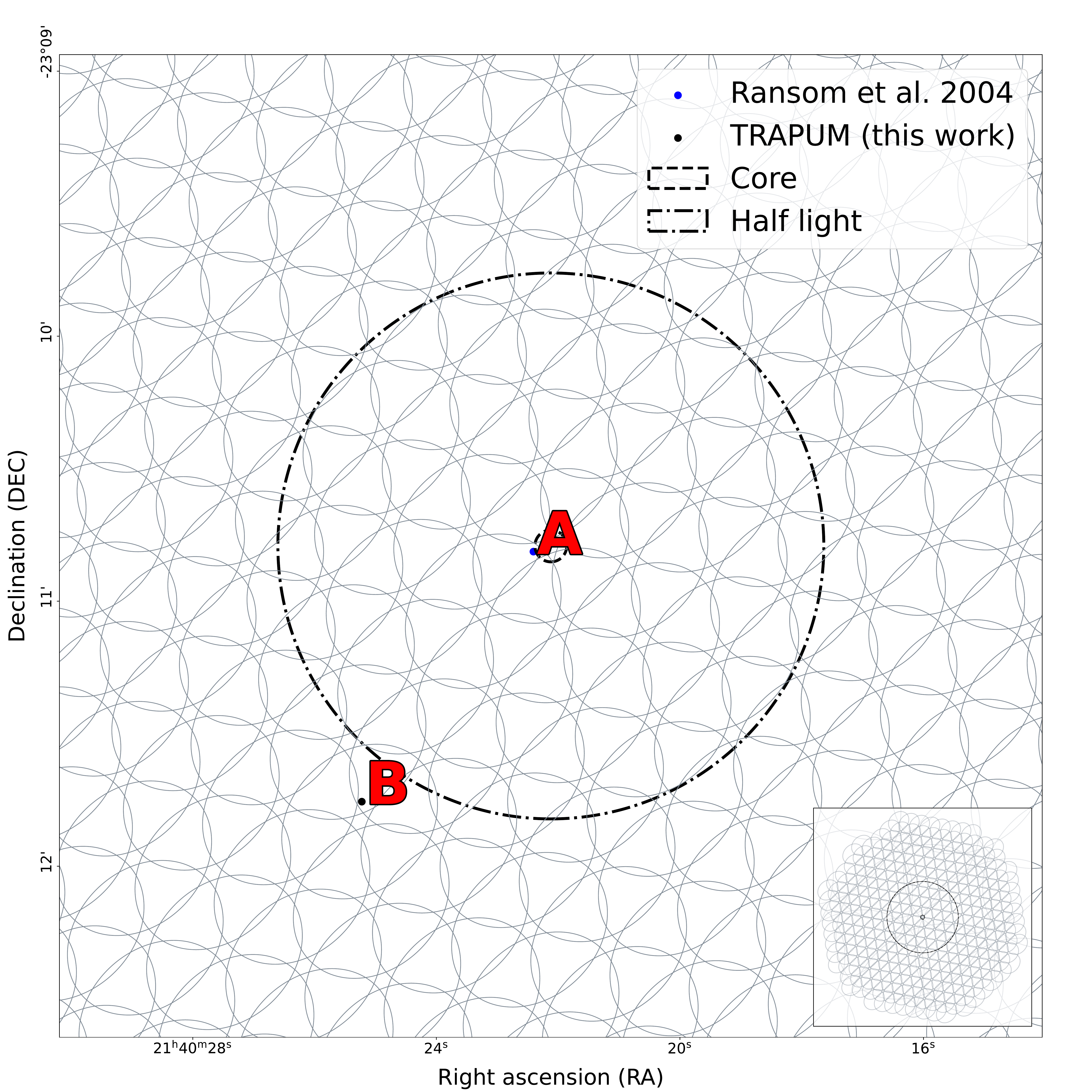}
    \caption{Beam tiling pattern at the start of the TRAPUM observation of M30 recorded in UHF band on 2021 January 22. Boundary of the ellipse plotted here corresponds to 50 \% of the maximum power level. The beam pattern was generated using \textsc{Mosaic}. The boundaries of the core and half-light radius have been obtained from the \citet{Harris2010} catalog which are marked as dashed and dash-dotted lines respectively. Overlaid on the plots in red are the two known pulsars within this cluster both of which were reported in \citet{2004ApJ...604..328R}. While M30A was discovered and regularly monitored since then, M30B remained undetected since the discovery detection which has now finally been confirmed, localized and solved using TRAPUM. The inset shows a zoomed out version of the same plot.}
    \label{fig:tiling diagram M30}
\end{figure*}

The full width at half maximum (FWHM) of the GBT beam at 20 cm is about \SI{9}{\arcminute}. When M30B was initially discovered, its position within the GBT telescope beam and by extension within the cluster was not yet known. The position of a pulsar with respect to the cluster centre gives us important clues regarding the evolutionary history of the system. As mentioned in \S \ref{sec:observation}, MeerKAT observations of this cluster made with the TRAPUM backend typically consist of over 270 synthesized beams (hereafter referred to as coherent beams) on the sky. These beams are much smaller than the typical sky area observed by a single-dish radio telescope at the same frequency. For example, the size\footnote{Defined here with an elliptical fit at 50-percent of the power level} of the semi-major and semi-minor axes of the coherent beam during the beginning of the January 22nd UHF band were 27.3 and 21.9 arcseconds respectively\footnote{The size of the coherent beam depends on the configuration of the antennas used, on the central frequency of the observation and on the source elevation}. Therefore, a detection in one of the beams already tells us that the pulsar is within or very near that beam. This position measurement can be improved further if detections in multiple beams are available which was the case with M30B where we detected the pulsar in five neighbouring beams in the 2021 January 22nd observation. We used the \textsc{SeeKAT} multibeam localiser software\footnote{\url{https://github.com/BezuidenhoutMC/SeeKAT}} (Bezuidenhout et al. submitted.) to improve our estimate for the position of this pulsar. \textsc{SeeKAT} takes in the position of the coherent beam, the detected S/N within it and the beam point-spread function (PSF) calculated using the software \textsc{Mosaic} and performs a maximum-likelihood analysis to get a better estimate of the pulsar's position. In Figure \ref{fig:tiling diagram M30}, we show the known radio timing position of M30A and the recently localized M30B's position within the cluster along with the TRAPUM beam tiling pattern for the beginning of the observation taken on 2021 January 22. M30B is located \SI{1.2(1)}{\arcminute} away from the cluster centre and just outside the half-light radius. The consequences of this in terms of the evolutionary history of M30B is discussed in \S \ref{sec:system origin}.

\subsection{Preliminary Orbital analysis}
Given our new detections from \textsc{TRAPUM} and \textsc{MeerTIME} observations, the first step was to obtain an accurate orbital solution for M30B which requires the identification of its orbital period $P_b$. In order to do this, we extracted the barycentric epoch and spin-period corresponding to each detection. We then used a modified version\footnote{\url{https://github.com/mcbernadich/CandyCracker}} of the \citet{2008MNRAS.387..273B} roughness algorithm to get an initial estimate of the orbital period of this system. Roughness $\phi$ is calculated by folding the data with a number of trial $P_b$ values. For each iteration, we then obtain data for the observed spin-period versus orbital phase. From this, we compute the summation of the squared differences of the observed spin period between adjacent values of $\phi$. The idea here is that the roughness should be minimum for the correct value of $P_b$. Building on the original algorithm, we added a variance measure to obtain a more robust measurement of $P_b$.  Using this method, we estimated the $P_{\rm b}$ of M30B to be 6.215 days. This estimate was then used as an initial guess to build a first-pass orbital ephemeris using the program \textsc{fitorbit}\footnote{\url{https://github.com/vivekvenkris/fitorbit}}.

\subsection{Pulsar timing}

The next step was to improve this orbital ephemeris through a process known as pulsar timing.  We started by folding all our observations using the \textsc{dspsr}\footnote{\url{https://dspsr.sourceforge.net/}} software package \citep{2011PASA...28....1V} modulo the predicted spin-period from the orbital ephemeris obtained from \textsc{fitorbit}. Then we formed a stable integrated pulse profile by summing the data in frequency, time, polarization. We then cross-correlated this high-S/N pulse profile with an analytic template to extract times of arrival (TOAs) at our telescope site for a particular rotational phase of the pulsar. These preprocessing steps were done using standard routines from the \textsc{PSRCHIVE}\footnote{\url{https://psrchive.sourceforge.net/}} \citep{2012AR&T....9..237V} software suite.

Using TOAs derived from our MeerKAT detections, we were able to determine the orbital parameters of M30B using the software \textsc{tempo}\footnote{\url{https://tempo.sourceforge.net/}} and the theory-independent ``DD'' model \citep{1986AIHS...44..263D}. To get a better estimate of DM, we extracted TOAs per frequency channel by summing each observation in time and polarization and averaging our data across frequency from 256 to 4 frequency channels. These TOAs were then used to fit for DM by keeping all other orbital and spin parameters fixed using \textsc{tempo}. Our ephemeris was precise enough to fold and detect the pulsar in the 2021 and 2022 UHF observations (obs id: 03U - 08U). However, since this ephemeris is not a phase-connected solution, a \textit{priori} we do not yet know the rotation count between groups of TOAs of different observations. This is estimated by adding an arbitrary time-offset for each observation in the form of `JUMP' statements between each set of locally connected TOAs, and then attempting to establish the rotation counts between close sets of observations, as described in detail in \S 3 of \cite{2018MNRAS.476.4794F}. Given the general sparsity of the detections, especially between 2001 and 2022, we could not connect all observations this way, so we used \textsc{Dracula} (described in \S 4 of \citealt{2018MNRAS.476.4794F}) to automatically identify the unknown number of rotations between observations. However, even with this algorithm, we cannot yet determine a unique phase-coherent timing solution for all the data of the pulsar, as the number of observations is still too small for that.

\begin{figure*}[ht!]
    \centering
    \includegraphics[width=1\textwidth]{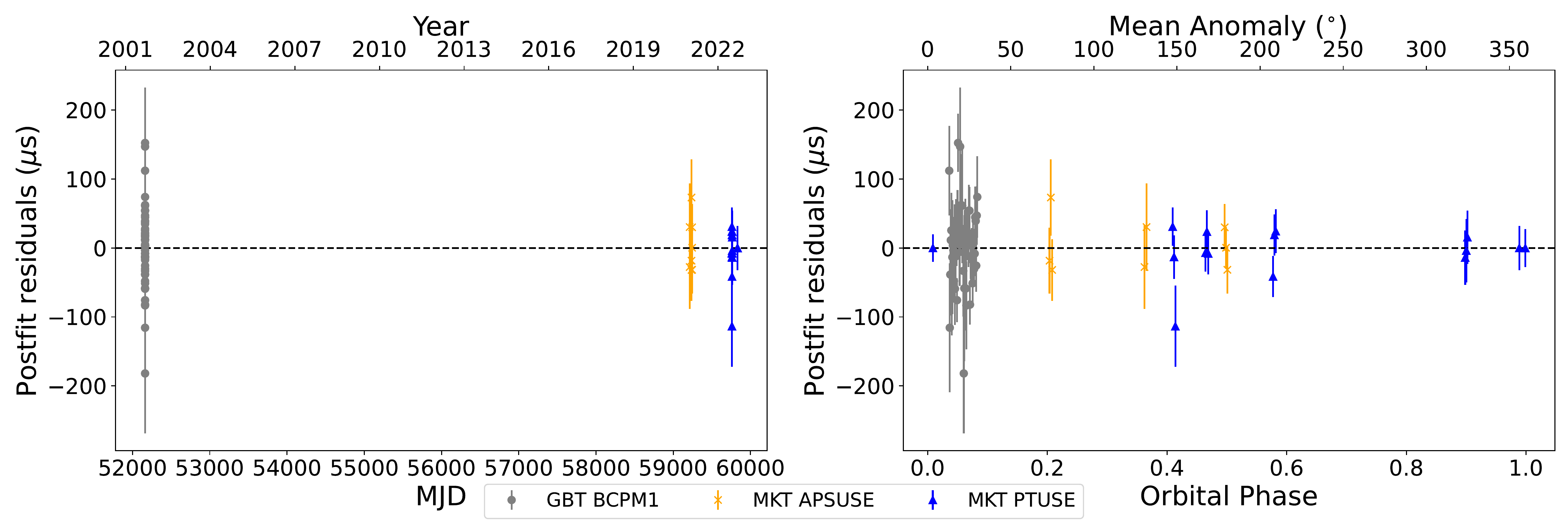}
    \caption{Postfit timing residuals of PSR~J2140$-$2311B as a function of time and orbital phase using the DD Binary model. Colors indicate different telescopes and backends used to obtain the TOAs.}
    \label{fig:post fit residuals M30B ddmodel}
\end{figure*}

However, we obtained a good orbital solution (see Table~\ref{tab:timing_solution_M30B_DD.par}, $\chi^2 = 1.36$) that can fold all the observations, including the early 2001 observation, and another observation taken during periastron on 2022 Sept (obs id: 09U). Additionally, all the TOAs from the 4-day orbital campaign (obs id: 05U-08U) made in 2022 are phase connected; this provides an important contribution to the precision of our solution. The post-fit residuals for this model can be found in Figure \ref{fig:post fit residuals M30B ddmodel} with the arbitrary time offsets subtracted. The flat residuals indicate that the model provides, within its current limitations, a good description of the timing. We were also able to recover the pulsar in one of our \textsc{MeerKAT} L-band observations (obs id: 02L) which was undetected by our blind searches. However, even with the aid of this ephemeris, we still could not recover the signal in any of the old GBT observations besides the discovery observation, nor in two 30-minute observations of M30 at a frequency of 400 MHz taken with the Giant Metrewave Radio Telescope (GMRT) in India as part of the GC survey presented by \cite{2022A&A...664A..54G}.

These non-detections can be explained because we do not yet have a phase coherent timing solution, but also because of the difference of sensitivity of the MeerKAT, GBT and GMRT observations. The survey sensitivity of TRAPUM GC observations for NGC 1851 have been reported previously in \S 2.1 of \citet{2022A&A...664A..27R} using the modified radiometer equation \citep{1985ApJ...294L..25D}. Adjusting these numbers for the values reported in table \ref{tab:list_observations m30b} for TRAPUM observations of M30, we get a minimum detectable flux density of $14.1\, \rm \mu Jy$ at L-BAND and $19.9\, \rm \mu Jy$ at the UHF band compared to $109.3\, \rm \mu Jy$ for the 400 MHz GMRT observation of M30 reported in \citet{2022A&A...664A..54G}. \citet{2004ApJ...604..328R} reported that the GBT observations of M30 were sensitive to normal millisecond pulsars in the range of $\sim 50 - 100 \,\rm \mu Jy$.  

\begin{table}
\caption{Timing parameters for PSR~J2140$-$2311B (M30B) presented in Barycentric Dynamical Time (TDB). For right ascension and declination, we report the values that correspond to the maximum likelihood reported by \textsc{SeeKAT} along with their one-sigma uncertainty. Both the position and DM have been kept fixed for our timing analysis given the sparsity of our detections.}
\begin{center}{\scriptsize
\setlength{\tabcolsep}{6pt}
\renewcommand{\arraystretch}{1.1}
\begin{tabular}{l c}
\hline
Pulsar  &   J2140$-$2311B                                                            \\
\hline\hline
Fitting program \dotfill & TEMPO \\
Time Units                                                            \dotfill &   TDB \\
Terrestrial Time Standard                                             \dotfill &   UTC(NIST)                                                              \\
Solar System Ephemeris                                                \dotfill &   DE440                                                                  \\
Right Ascension, $\alpha$ (J2000)                                     \dotfill &   21:40:25.2(3)                                                                      \\
Declination, $\delta$ (J2000)                                         \dotfill &     $-$23:11:45(7)                                                                     \\

Spin Frequency, $f$ ($\rm Hz$)                                        \dotfill &   76.9833055(6)                                                          \\

1st Spin Frequency derivative, $\dot{f}$ ($\rm Hz \, s^{-1}$)                                        \dotfill &  3(15)$\rm \times 10^{-15}$                                                           \\

Reference Epoch (MJD)                                                 \dotfill &   59763.520924                                                           \\
Start of Timing Data (MJD)                                            \dotfill &   52161.993                                                              \\
End of Timing Data (MJD)                                              \dotfill &   59831.942                                                              \\
Dispersion Measure, DM (pc cm$^{-3}$)                                 \dotfill &   25.063(3)                                                             \\
                                                                    
Number of TOAs                                                        \dotfill &   73                                                                     \\
Residuals RMS ($\mu$s)                                                \dotfill &   39.42                                                                  \\
\hline
\multicolumn{2}{c}{Binary Parameters}  \\
\hline\hline
Binary Model                                                          \dotfill &   DD                                                                     \\
Projected Semi-major Axis, $x_{\rm p}$ (lt-s)                         \dotfill &   19.5222(7)                                                             \\
Orbital Eccentricity, $e$                                             \dotfill &   0.87938(2)                                                            \\
Longitude of periastron, $\omega$ (deg)                               \dotfill &   160.8007(4)                                                            \\
Epoch of periastron Passage, $T_0$ (MJD)                           \dotfill &   59763.520649(6)                                                        \\
Rate of periastron advance, $\dot{\omega}$ (deg/yr)                   \dotfill &   0.078(2)                                                 \\
Orbital Period, $P_b$ (days)                                          \dotfill &   6.21565400(6)                                                          \\

\hline
\multicolumn{2}{c}{Derived Parameters}  \\
\hline\hline
Galactic longitude, $l$ ($^\circ$) \dotfill & 27.161(1) \\
Galactic latitude, $b$ ($^\circ$) \dotfill & $-$46.851(2) \\
Total system Mass, $M_{\rm TOT}$ (M$_\odot$)                                           \dotfill &   2.53(8)              \\
Companion mass, $M_{\rm c}$ (${\rm M}_\odot$)            \dotfill &   $\ge$ 1.10                                                                      \\
                                                                     
Pulsar mass, $M_{\rm p}$ (${\rm M}_\odot$)             \dotfill &   $\le$ 1.43                                                                      \\
Spin Period, $P$ (s)                                                  \dotfill &   0.01298982933(2)                                         \\
1st Spin Period derivative, $\dot{P}$ (s s$^{-1}$)                    \dotfill &   $-$6(25)$\times 10^{-19}$                                            \\
Mass Function, $f(M_{\rm p})$ (${\rm M}_\odot$)                       \dotfill &   0.2067                                                          \\

Total offset from GC center, $\theta_\perp$ (arcmin)                  \dotfill &   1.2(1)                                                                     \\

\hline
\end{tabular} }
\end{center}
\label{tab:timing_solution_M30B_DD.par}
\end{table}

\begin{figure*}[ht!]
    \centering
    \includegraphics[width=.8\textwidth]{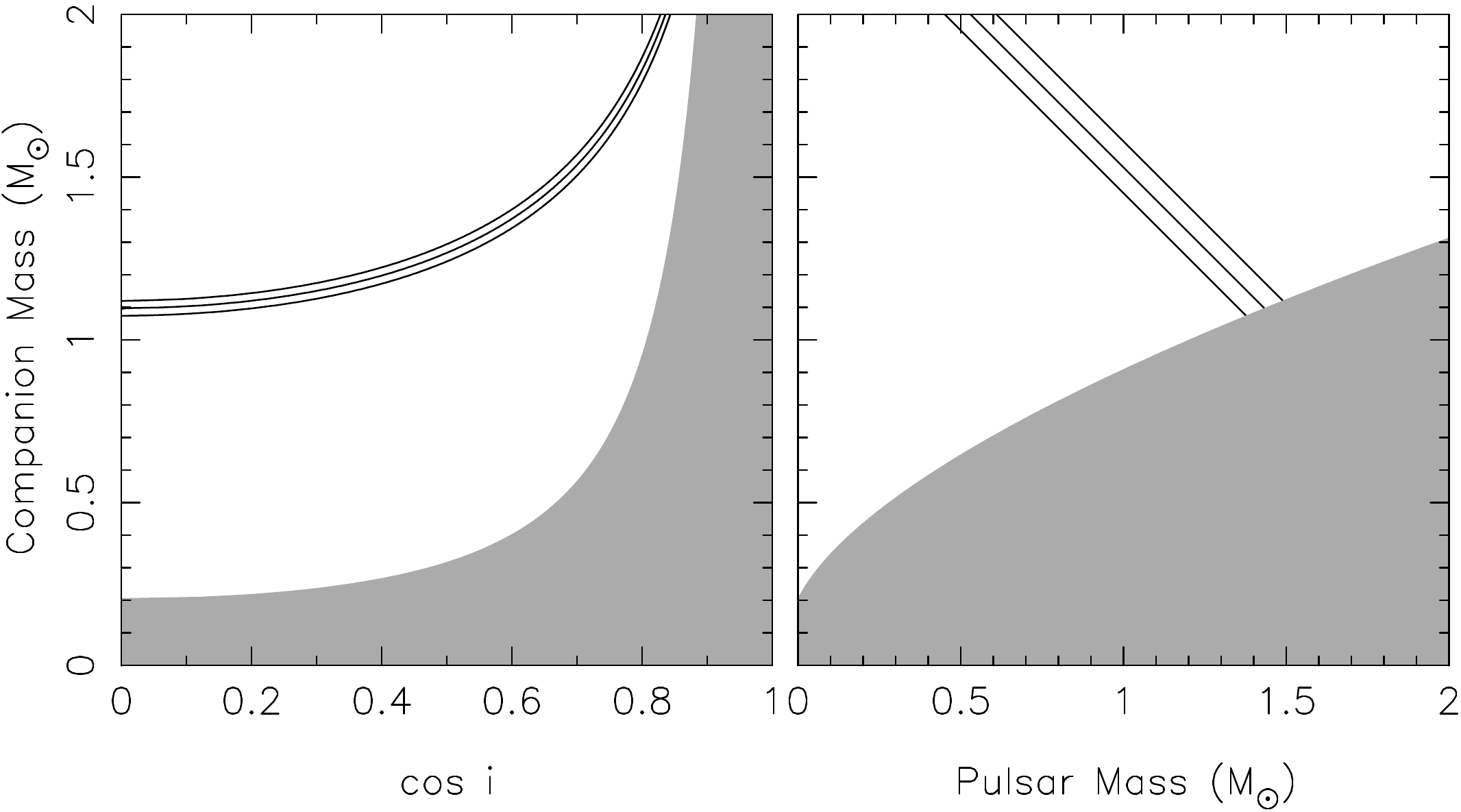}
    \caption{Mass-mass diagram of M30B assuming GR obtained from pulsar timing. In the left plot, we show the mass of the companion $M_{\rm c}$ versus the cosine of the inclination angle. The gray region here is excluded by the requirement that $M_{\rm p} > 0$. On the right we present $M_{\rm c}$ versus the mass of the pulsar $M_{\rm p}$. The gray region is excluded by the mass function and orbital geometry (i.e $\sin{i} \leq 1$). The black triplet of lines shows the regions consistent with the measurement of the advance of periastron $\dot{\omega}$ along with their 1 $\sigma$ uncertainty. As we can see, from these measurements we can derive an upper limit for $M_{\rm p}$ and a lower limit for $M_{\rm c}$.}
    \label{fig:mass mass diagram M30B}
\end{figure*}

An important parameter in this orbital solution is the system's rate of advance of periastron, which we measure to be $\dot{\omega} = 0.078 \pm 0.002\, \rm deg \, yr^{-1}$. The precision of this measurement greatly benefits from the inclusion of the 2001 TOAs, even with the fit of an arbitrary time offset.
Using the DDGR model \citep{1987grg..conf..209T, 1989ApJ...345..434T}, which assumes that general relativity (GR) accounts for the relativistic effects observed in the timing, the total mass of the system derived from $\dot{\omega}$ is $M_{\rm tot} = 2.53 \pm 0.08$ M$_{\odot}$. The mass-mass diagram assuming GR along with our measurement of $\dot{\omega}$ is shown in Figure \ref{fig:mass mass diagram M30B}. In this Figure, we can see that combining the total mass measurement with the
constraint from the mass function, we can estimate a minimum companion mass of 1.10 M$_{\odot}$ and a maximum pulsar mass of 1.43 M$_{\odot}$.

Note that, although our measurement of the first-spin frequency derivative $\dot{f}$ is not significant, it is important to fit for this parameter in order to obtain a realistic uncertainty for $\dot{\omega}$. Indeed, if we don't fit for $\dot{f}$, the uncertainty of $\dot{\omega}$ will be one order of magnitude smaller. Assuming the small $\dot{f}$ typical of MSPs is not warranted because it could have a significant contribution from the system's acceleration in the cluster.

% PF: There was a bit here about future timing that I merged with what is now 3.6

\subsection{System origin}\label{sec:system origin}
Most binary MSPs in our Galaxy are in highly circularised orbits with a low-mass He-WD or a non-degenerate or semi-degenerate ultra-light companion (M$_c$ $\leq$ 0.08 M$_{\odot}$, \citealt{2012MNRAS.425.1601T} and references within). These systems are highly recycled ($P \leq \SI{10}{\milli \second}$) due to the long mass transfer phase which transfers matter and angular momentum from the companion to the pulsar, spinning it up to very rapid rates and reducing the magnetic field strengths of the NS \citep{1991PhR...203....1B, 2002JApA...23...67B, 2006csxs.book..623T}. Intermediate spin-period pulsars ($\SI{10} \leq \;  P \leq \SI{20}{\milli \second}$, \citealt{2001ApJ...548L.187C}) tend to have massive CO or ONeMg WD companions (e.g. PSR~J1802$-$2104, \citealt{2010ApJ...711..764F}), but their orbits still have very low eccentricities.

If the companion star is massive enough to undergo its own supernova (SN) explosion and if the binary orbit survives, then a double neutron star system (DNS, see \citealt{2017ApJ...846..170T} for a review) will form. These systems tend to be mildly recycled ($P \geq \SI{17}{\milli \second}$, see \citealt{2018ApJ...854L..22S}) as their massive companions do not live for long, halting the recycling process earlier. Unlike systems with WD companions, they have highly eccentric orbits because of the kick and mass loss from the second SN \citep{1995MNRAS.274..461B,2017ApJ...846..170T}.

Given its high orbital eccentricity, M30B could in principle have formed like the DNSs in the Galactic disk. The total mass is similar to that of the lightest known DNS systems (like PSRs~J1411+2551 and J1946+2052, \citealt{2017ApJ...851L..29M,2018ApJ...854L..22S}). However, this scenario is unlikely given the spin period of M30B --- $P \sim \SI{12.98}{\milli \second}$, faster than for any pulsars in a DNS seen to date in the Galaxy. Furthermore, there are currently no massive stars in GCs that would provide a second SN; the last time these existed in GCs was more than 10 Gyr ago. Thus, if M30B was such a primordial DNS, its characteristic age ($\tau_{\rm c}$) would have to be at least $\sim 10 \, \rm Gyr$. We note, though, that given the large $\gamma$ of the host cluster it is very unlikely that the system would still resemble its original configuration.

The spin period of M30B is more compatible with a relatively massive CO or ONeMg WD companion. In the Galactic disk, these systems have low orbital eccentricities, but given the large number of stellar encounters in GCs, the orbital eccentricity could have been greatly increased \citep{1992RSPTA.341...39P}. Thus, it is possible that the companion is a massive WD star whose progenitor recycled the pulsar; in this case we should also expect $\tau_{\rm c} \sim 10 \, \rm Gyr$. 

A more likely hypothesis is that M30B is a result of an exchange encounter, where the lighter star that recycled the pulsar was ejected during the binary's chaotic encounter with a more massive degenerate star, which is the current companion. Given the random dynamics of such an encounter, we cannot decide on the nature of the companion - either a massive WD or a NS --- based on arguments from stellar evolution (although massive white dwarfs are generally more likely given their larger abundance in the cluster, see e.g., \citealt{2019ApJ...877..122Y,2021ApJ...917...28K}). %This expectation is corroborated by the distribution of companion masses measured in these eccentric binaries. 

Several eccentric MSP binaries with massive companions recently been found in core-collapsed GCs (which have the highest $\gamma$ values: PSR~J1807$-$2500B in NGC~6544 \citealt{2012ApJ...745..109L}, PSR~J1835$-$3259A in NGC 6652 \citealt{2015ApJ...807L..23D} and PSR~J1823$-$3021G in NGC 6624 \citealt{2021MNRAS.504.1407R}). These are trought to be exchange products because their spin periods are very small compared to binary pulsars in the Galactic disk with similarly massive companions. In all core-collapsed GCs, these systems represent approximately 1/3 of the known population of binary radio pulsars. For this calculation, we have adopted the definition of core-collapsed GC from the Harris catalog (\citealt{1996yCat.7195....0H}, 2010 revision) and only considered binary pulsars with well measured orbits. We then assume that pulsars in an eccentric orbit ($e > 0.1$) orbiting a massive companion ($m_c > 0.38 \; \rm M_{\odot}$) are likely to be the result of exchange products. Using this definition and the updated numbers from the `GC Pulsar catalog', we find that 5 out of the 14 ($\sim$ 35.7 \%) binary pulsars in core-collapsed GCs are likely to be exchange products; in non core-collapsed GCs only 10 out of 109 ($\sim$ 9.2\%) binary pulsars fulfill these criteria. The location of M30B in a core-collapsed GC is thus an indication that it likely originated in an exchange encounter. Another clue is its high eccentricity: Among secondary exchange products, M30B has the third-highest orbital eccentricity after PSR~J1835$-$3259A ($e \sim 0.96$, \citealt{2015ApJ...807L..23D}) and PSR~J0514$-$4002A ($e \sim 0.88$, \citealt{2007ApJ...662.1177F, 2019MNRAS.490.3860R}).

When exchange encounters are very frequent, they might even happen during the LMXB phase. In this case, the recycling process is truncated, resulting in a partially recycled pulsar that still has a relatively high magnetic field and will therefore appear young. Unlike MSPs, such mildly recycled pulsars spin down fast, which means that their LMXB disruption must be recent. This explains why slow, apparently young pulsars in GCs are overwhelmingly found in high-$\gamma$ GCs and lie below the pulsar spin-up line \citep{2014A&A...561A..11V,2022MNRAS.513.2292A}, although in this regard we must keep in mind that there are alternative explanations for the formation of apparently young pulsars in GCs (e.g. \citealt{2008MNRAS.386..553I}).

The recent disruption of a LMXB by a massive degenerate intruder that then becomes the pulsar's companion is a likely explanation for the parameters of PSR~B2127+11C ($P = 30.5 \, \rm ms$ and $\tau_{\rm c} \sim 0.1 \, \rm Gyr$), a binary pulsar located in the core-collapsed GC M15. Although this system superficially resembles a Galactic DNS (see \citealt{2019ApJ...880L...8A}), its characteristic age - less than 1\% of the age of the GC - is too small for it to be a DNS formed from the primordial population of massive stars of M15 \citep{1991ApJ...374L..41P}. If the recent disruption of a LMXB is the explanation for the relatively slow spin of M30B, then we might also expect the pulsar to have a characteristic age much smaller than the age of the M30 GC itself.

One clue that suggests that PSR~B2127+11C was recently involved in an exchange encounter is its large distance (\SI{0.94}{\arcminute}) from the centre of M15 \citep{1991ApJ...374L..41P,2006ApJ...644L.113J}. Normally, mass segregation causes the pulsar population in dense GCs to be very centrally condensed, with most pulsars within, or near their cores (e.g., \citealt{2017MNRAS.471..857F,2017ApJ...845..148P,2018MNRAS.481..627A}). This is the case for all other pulsars in M15 \citep{1993PhDT.........2A} and for M30A \citep{2004ApJ...604..328R}.
The large distance of PSR~B2127+11C from the center of M15 is possibly the result of the recoil caused by the ejection of the previous light companion that partially recycled the pulsar;  the time elapsed since the recoil ($\sim 0.1 \rm Gyr$, if it is the same event that disrupted the LMXB) is presumably too short for the system to have migrated back to the centre of the GC via dynamical friction. Interestingly, M30B is located \SI{1.2(1)}{\arcminute} from the cluster centre and just outside its half-light radius, which again suggests a recent exchange interaction. Overall, pulsars with such large distances from the centre (in core radii) occur more often in high-$\gamma$ GCs \citep{2014A&A...561A..11V}.

\subsection{Prospects}
% PF: small paragraph moved in here from section 3.4
 We plan to continue timing M30B as part of the MeerTime GC pulsar timing programme. These observations are necessary in order to fully connect all the MeerKAT observations (possibly all the way back to 2001) and shed light into the nature of the M30B system. A phase-coherent timing solution will yield much improved astrometric, spin and orbital parameters. The spin and orbital period derivatives will be extremely important, because their measurement will allow the determination of $\tau_c$, which will be crucial for distinguishing between a primordial binary with a massive companion or a more recent exchange product. This will also greatly improve all orbital parameters, especially $P_b$ and $\dot{\omega}$.
 
 The following step will be to determine the individual masses via the detection of additional relativistic effects. Dense orbital campaigns in the near future might lead to the detection of the Shapiro delay \citep{1964PhRvL..13..789S}: for an inclination angle of $60^{\circ}$, the $h_3$ parameter \citep{2010MNRAS.409..199F} would be $1.2 \mu s$. This can be detected with 2-$\sigma$ significance with 4000 ToAs with the current timing precision. The parameter will be larger (and therefore detected with higher significance) for higher inclinations, another possibility is to somehow improve the timing precision. If the orbital inclination is low and/or we're unable to improve the timing precision, then we will have to wait to detect the Einstein delay. However, the longitude of periastron $\omega$ of M30B is not optimal for this goal (see detailed discussion in \citealt{2019MNRAS.490.3860R}), so if the Shapiro delay is not detectable, measuring the component masses with the Einstein delay will take several decades.

\section{Conclusion}
In this letter, we presented the confirmation of M30B with the first set of new detections of this pulsar since its discovery in 2001. We found that the pulsar can be reliably detected with the MeerKAT UHF receivers; this has finally allowed, 20 years after the discovery, a detailed characterization of this system: it is located \SI{1.2(1)}{\arcminute} from the cluster centre and the pulsar is in a highly eccentric ($e = 0.879$) orbit around a companion that could either be a massive WD or a NS. We also measured the rate of periastron advance, which indicates a total system mass consistent with that of the lightest known DNSs in our Galaxy \citep{2017ApJ...851L..29M,2018ApJ...854L..22S}\footnote{For a list of NS mass measurements, see \url{https://www3.mpifr-bonn.mpg.de/staff/pfreire/NS_masses.html}}. M30B was likely formed as the result of a secondary exchange encounter, similar systems have been observed in other GCs with very dense cores. Further timing observations are necessary to obtain a phase-connected timing solution for this pulsar, which would yield much improved astrometric, spin and orbital parameters. Continued timing might result in the detection of additional relativistic effects and the determination of the individual masses of the components. The characterization of M30B is a demonstration of the unrivalled sensitivity of MeerKAT for radio sources in the Southern celestial hemisphere.

\begin{acknowledgments}
The MeerKAT telescope is operated by the South African Radio Astronomy Observatory, which is a facility of the National Research Foundation, an agency of the Department of Science and Innovation. SARAO acknowledges the ongoing advice and calibration of GPS systems by the National Metrology Institute of South Africa (NMISA) and the time space reference systems department of the Paris Observatory.
PTUSE was developed with support from  the Australian SKA Office and Swinburne University of Technology. MeerTime data is housed on the OzSTAR supercomputer at Swinburne University of Technology. The OzSTAR program receives funding in part from the Astronomy National Collaborative Research Infrastructure Strategy (NCRIS) allocation provided by the Australian Government.
The authors also acknowledge MPIfR funding to contribute to MeerTime infrastructure.
TRAPUM observations used the FBFUSE and APSUSE computing clusters for data acquisition, storage and analysis. These clusters were funded and installed by the Max-Planck-Institut für Radioastronomie and the Max-Planck-Gesellschaft. The National Radio Astronomy Observatory is a facility of the National Science Foundation operated under cooperative agreement by Associated Universities, Inc. The Green Bank Observatory is a facility of the National Science Foundation operated under cooperative agreement by Associated Universities, Inc. 
VB, AR, EDB, FA, DJC, WC, PCCF, TG, MK, PVP and VVK  acknowledge continuing valuable support from the Max-Planck Society. SMR is a CIFAR Fellow and is supported by the NSF Physics Frontiers Center awards 1430284 and 2020265.
This work is supported by the Max-Planck Society as part of the "LEGACY" collaboration on low-frequency gravitational wave astronomy.
AR and AP gratefully acknowledge financial support by the research grant ``iPeska'' (P.I. Andrea Possenti) funded under the INAF national call Prin-SKA/CTA approved with the Presidential Decree 70/2016. AR and AP also acknowledge support from the Ministero degli Affari Esteri e della Cooperazione Internazionale - Direzione Generale per la Promozione del Sistema Paese - Progetto di Grande Rilevanza ZA18GR02.
BWS acknowledges funding from the European Research Council (ERC) under the European Union’s Horizon 2020 research and innovation programme (grant agreement No. 694745). Pulsar research at UBC is supported by an NSERC Discovery Grant and by the Canadian Institute for Advanced Research.
\end{acknowledgments}

\bibliography{M30B_APJ_letter}{}
\bibliographystyle{aasjournal}

\end{document}